\documentclass[lettersize,journal]{IEEEtran}
\usepackage{amsmath,amsfonts}
\usepackage{algorithm}
\usepackage{array}
\usepackage[caption=false,font=normalsize,labelfont=sf,textfont=sf]{subfig}
\usepackage{textcomp}
\usepackage{stfloats}
\usepackage{url}
\usepackage{graphicx}
\usepackage{cite}
\usepackage{booktabs} 
\usepackage[para]{threeparttable} 

\begin{document}

\title{Electromagnetic Properties of Aluminum-based Bilayers for Kinetic Inductance Detectors}

\author{G.~Wang, P.~S.~Barry, T.~Cecil, C.~L.~Chang, J.~Li, M.~Lisovenko, V.~Novosad, Z.~Pan, \\
V.~G.~Yefremenko, and J. Zhang
\thanks{Manuscript receipt and acceptance dates will be inserted here. Work at Argonne, including use of the Center for Nanoscale Materials, an Office of Science user facility, was supported by the U.S. Department of Energy, Office of Science, Office of Basic Energy Sciences and Office of High Energy Physics, under Contract No. DE-AC02-06CH11357.}
\thanks{G. Wang is with Argonne National Laboratory, Lemont, IL 60439 USA (\textit{Corresponding author}. Email:gwang@anl.gov).}%
\thanks{P. S. Barry is with Cardiff University, (Email: barryp2@cardiff.ac.uk).}%
\thanks{T. Cecil, J. Li, M. Lisovenko, V. Novosad, Z. Pan, V. G. Yefremenko, and J. Zhang are with Argonne National Laboratory, Lemont, IL 60439 USA (Email: cecil@anl.gov; juliang.li@anl.gov; mlisovenko@anl.gov; novosad@anl.gov; panz@anl.gov; yefremenko@anl.gov; jianjie.zhang@anl.gov).}%
\thanks{C. L. Chang is with Argonne National Laboratory, Lemont, IL 60439 USA and the University of Chicago, Chicago, IL 60637 USA (Email: clchang@kicp.uchicago.edu).}%
}

\markboth{3EPo2D-06}%
{Shell \MakeLowercase{}}


\maketitle

\begin{abstract}
The complex conductivity of a superconducting thin film is related to the quasiparticle density, which depends on the physical temperature and can also be modified by external pair breaking with photons and phonons. This relationship forms the underlying operating principle of Kinetic Inductance Detectors (KIDs), where the detection threshold is governed by the superconducting energy gap. We investigate the electromagnetic properties of thin-film aluminum that is proximitized with either a normal metal layer of copper or a superconducting layer with a lower $T_C$, such as iridium, in order to extend the operating range of KIDs. 
Using the Usadel equations along with the Nam expressions for complex conductivity, we calculate the density of states and the complex conductivity of the resulting bilayers to understand the dependence of the pair breaking threshold, surface impedance, and intrinsic quality factor of superconducting bilayers on the relative film thicknesses. 
The calculations and analyses provide theoretical insights in designing aluminum-based bilayer kinetic inductance detectors for detection of microwave photons and athermal phonons at the frequencies well below the pair breaking threshold of a pure aluminum film.
\end{abstract}


\section{Introduction}
\IEEEPARstart{S}{uperconducting} microwave Kinetic Inductance Detectors (KIDs)~\cite{Zmuidzinas:12} are an attractive detector technology owing to the ability to read out a large number of pixels without the need for ancillary cryogenic multiplexing components.
KIDs have been developed to measure photons~\cite{Austermann:18, Dober:14, Karkare:22, Meeker:18, Walter:20, Calvo:16} in astrophysics, athermal phonons~\cite{Moore:12, Cornell:14, Cardani:18, Colantoni:20} for dark matter and neutrinoless double beta decay searches, temperature changes~\cite{Miceli:14, Ulbricht:15, Wandui:20} caused by deposited energy of X-rays and long wavelength photon flux, and electric current or magnetic field~\cite{Vissers:15, Kher:217, Luomahaara:14}. 
For optimized performance of KIDs, superconducting bilayers or trilayers can be used to tune the pair breaking threshold~\cite{Vissers:13, Hu:20, Catalano:15, Aja:20, Sekine:14}, to increase the kinetic inductance fraction~\cite{Vissers:13, Hu:20, Catalano:15}, to increase photon energy readout efficiency~\cite{Zobrist:22}, and to enhance the current dependence of the kinetic inductance~\cite{Claassen:99, Ustavschikov:20}.

In this work, we investigate the electromagnetic properties of aluminum-based bilayers as a method to tune the pair breaking threshold for applications including line intensity mapping of carbon monoxide at high redshift~\cite{Mashian:15,Dizgah:22}, and searches for dark matter~\cite{Moore:12, Colantoni:20} and neutrinoless double beta decay~\cite{Cardani:18}. 
In particular, a lower pair breaking threshold extends the sensitivity of a detector to photons and phonons at lower frequencies. 

We start with the theory of proximity bilayers and methods to calculate density of states, complex conductivity, and surface impedance in Section II, present our calculation approaches and results for thin Cu/Al and Ir/Al bilayers in Section III, and conclude in Section IV. 

\section{Theory of a Proximity Bilayer}
\label{sec:theory}
We consider a bilayer, which consists of superconductor, 1, with a thickness of $d_1$ and a transition temperature of $T_{C1}$ on the left and a superconductor, 2, with a thickness of $d_2$ and a transition temperature of $T_{C2}$ on the right. 
In the dirty limit, where the electron mean free path is less than coherence length, the superconducting properties of the bilayer can be described with the Usadel equations~\cite{Usadel:70, Golubov:04, Golubov:95, Zhao:99, Brammertz:01, Zhao:17} using a function $\theta(x,E)$ known as the pairing angle, where $x$ is a position coordinate with $x = 0$ at the interface. 
The energy, $E$, is given by $E=j\hbar\omega_n$ with Matsubara frequency of $\omega_n=(2n+1)\pi k_BT / \hbar$ $(n=0$, $1$, $2$, \ldots$)$, where $\hbar$ is the reduced Planck constant, $k_B$ is the Boltzmann constant, and $T$ is temperature. 
The variable $\theta(x,E)$ is complex and ranges in magnitude from $0$ to $\pi/2$. 
The Usadel equations~\cite{Zhao:99, Brammertz:01, Zhao:17} are
\begin{equation}
\label{eqn:usadel}
\begin{aligned}
\frac{\hbar D_{1,2}}{2} \frac{\partial^2 \theta_{1,2} }{\partial x^2} -\hbar \omega_n \sin\theta_{1,2} + \Delta_{1,2} (x) \cos\theta_{1,2}=0,  
\end{aligned}
\end{equation}
where $D_{1,2}=\sigma_{1,2} / e^2N_{1,2}$ are the diffusivity of electrons in superconductor 1 and 2, $\sigma_{1,2}$ are the normal state conductivity, and $e$ is the electron charge. 
$N_{1,2}=3 \gamma_{1,2} / \pi^2 k_B^2 (1+\lambda_{1,2})$ are the density of states~\cite{John:84} at Fermi energy, where $\gamma_{1,2}$ are electronic specific heat coefficient, and $\lambda_{1,2}$ are electron phonon coupling constant.
The pair potentials, $\Delta_{1,2} (x)$, are determined by 
\begin{equation}
\label{eqn:dlog}
\begin{aligned}
\Delta_{1,2} \ln \Big( \frac{T}{T_{C1, C2}} \Big) = -2\pi k_B T \sum_{n\geq0} \Big( \frac{\Delta_{1,2}}{\hbar \omega_n} - \sin\theta_{1,2} \Big).  
\end{aligned}
\end{equation}

At the open boundaries facing either vacuum or a dielectric, no current flows~\cite{Kuprianov:88, Fominov:01, Zhu:09}, therefore 
\begin{equation}
\label{eqn:BC1}
\begin{aligned}
\frac{\partial \theta_1}{\partial x} = \frac{\partial \theta_2}{\partial x} =0.  
\end{aligned}
\end{equation}

At the interface of the two superconductors, the current across the interface is conserved~\cite{Kuprianov:88, Fominov:01, Zhu:09}, therefore
\begin{equation}
\label{eqn:BC2}
\begin{aligned}
\sigma_1 \frac{\partial \theta_1}{\partial x} = \sigma_2 \frac{\partial \theta_2}{\partial x} = \frac{G_{int}}{A} \sin(\theta_2 - \theta_1),  
\end{aligned}
\end{equation}
where $A$ is the area of the interface between the two superconductors. $G_{int} = 2 t N_{ch} G_K$ is the Landauer conductance across the interface~\cite{Martinis:00}, where $t$ is a free parameter describing electron transmission coefficient across the interface, $N_{ch}=A/(\lambda_f/2)^2$ is the number of conductance channels, and $\lambda_f$ is the Fermi wavelength in the superconductor that has fewer conductance channels. $G_K=e^2/h$ is the conductance quantum and $h$ is the Planck constant. 

With the solutions of $\theta(x,E)$ from Eqs.~\ref{eqn:usadel} to~\ref{eqn:BC2}, we can calculate the single-spin density of electron states $N(x,E)=N_{1,2} n(x,E)$, where $n(x,E)=$ Re$[\cos{\theta(x,E)}]$ (see Eq.~\ref{eqn:n}), as well as the pair breaking threshold $2\Delta_g$. Since the energy dependence of $n(x,E)$ is sharp (similar to the BCS density of states), we define $\Delta_g=E$ when $n(x,E)=0.001$.

The complex conductivity of a superconductor can be written as $\sigma=\sigma_1-j\sigma_2$. 
Normalized with the normal state conductivity $\sigma_N$, the Nam’s formulation of complex conductivity of strong-coupling and impure superconductors~\cite{Nam:67} is 
\begin{equation}
\label{eqn:sigma1}
\begin{aligned}
\frac{\sigma_1}{\sigma_N}=& \frac{1}{\hbar\omega} \bigg\{ \int_{\Delta_g-\hbar\omega}^{-\Delta_g} g_1(1,2) \tanh \bigg( \frac{\hbar\omega+E}{2k_BT} \bigg) dE\\
            &+ \int_{\Delta_g}^{\infty} g_1(1,2)\bigg[ \tanh \bigg(\frac{\hbar\omega+E}{2k_BT} \bigg)\\ 
            &- \tanh \bigg(\frac{E}{2k_BT} \bigg) \bigg] dE \bigg\},
\end{aligned}
\end{equation}
\begin{equation}
\label{eqn:sigma2}
\begin{aligned}
\frac{\sigma_2}{\sigma_N}=& \frac{1}{\hbar\omega} \bigg\{ \int_{[\Delta_g-\hbar\omega,-\Delta_g]}^{\Delta_g} g_2(1,2) \tanh \bigg( \frac{\hbar\omega+E}{2k_BT} \bigg) dE\\
            &+ \int_{\Delta_g}^{\infty} \bigg[ g_2(1,2) \tanh \bigg(\frac{\hbar\omega+E}{2k_BT} \bigg)\\ 
            &+ g_2(2,1) \tanh \bigg(\frac{E}{2k_BT} \bigg) \bigg] dE \bigg\},
\end{aligned}
\end{equation}
where $\omega$ is the angular frequency of the excitation electromagnetic wave and $[\Delta_g-\hbar\omega,-\Delta_g]$ denotes that the algebraically larger of the two numbers is to be used. Here the functions $g_1$ and $g_2$ are coherence factors defined by 
\begin{equation}
\label{eqn:g112}
\begin{aligned} 
g_1 (1,2) = n(E)  n(E+\hbar \omega) + p(E)  p(E+\hbar \omega),
\end{aligned}
\end{equation}
\begin{equation}
\label{eqn:g212}
\begin{aligned} 
g_2 (1,2) = -\tilde{n}(E)  n(E+\hbar \omega) -\tilde{p}(E) p(E+\hbar \omega),
\end{aligned}
\end{equation}
where  $n(E)$, $\tilde{n}(E)$, $p(E)$, and $\tilde{p}(E)$ are defined by
\begin{equation}
\label{eqn:n}
\begin{aligned} 
n(E) + j \tilde{n}(E) = \cos\theta(x,E),
\end{aligned}
\end{equation}
\begin{equation}
\label{eqn:p}
\begin{aligned} 
p(E) + j \tilde{p}(E) = -1j* \sin\theta(x,E),
\end{aligned}
\end{equation}
where $\theta(x,E)$ comes from the solution of the Usadel equations. Both the real and imaginary parts of the complex conductivity are functions of location and energy. The upper limit of the integrations in Eqs.~\ref{eqn:sigma1} and~\ref{eqn:sigma2} is the Debye energy.

The surface impedance of a multilayer can be estimated by cascading the transmission matrices of all layers as formulated by Zhao et al.~\cite{Zhao:17}. In the case of a thin bilayer, we have
\begin{equation}
\label{eqn:matrix}
\begin{aligned} 
\begin{bmatrix}
v_S \\
i_S
\end{bmatrix}
=\prod_k
\begin{bmatrix}
1 & j \omega \mu_0 d_k \\
\sigma_k(\omega) d_k & 1 
\end{bmatrix}
\begin{bmatrix}
v_0 \\
i_0
\end{bmatrix},
\end{aligned}
\end{equation}
where $\sigma_k$ is the averaged complex conductivity of superconductor 1 or 2 of the thin bilayer, $\mu_0$ is the free space permeability, and $v_0 / i_0 = Z_0$ is the impedance of free space.  
The surface impedance is calculated as $Z_S = v_S / i_S$. Typically, the surface impedance can be written as
\begin{equation}
\label{eqn:Zs}
\begin{aligned} 
Z_S = R_S + j X_S,
\end{aligned}
\end{equation}
where $R_S$ represents resistive loss, and $X_S = \omega L_S$ is the reactance, where $L_S$ is known as surface inductance. We introduce a surface impedance quality factor~\cite{Zmuidzinas:12, Zhao:17} as
\begin{equation}
\label{eqn:Qs}
\begin{aligned} 
Q_S = X_S / R_S,
\end{aligned}
\end{equation}
which can be estimated through a measurement of the quality factor of a planar  superconducting resonator~\cite{Zmuidzinas:12}.

\section{Calculation Approaches and Results}
\label{sec:Results}
We numerically solve the Usadel Eqs. (Eqs.~\ref{eqn:usadel} and~\ref{eqn:dlog}) in a self-consistent procedure. 
Eq.~\ref{eqn:usadel} is linearized in matrix form using the finite difference method. 
We start with the trial pair potentials, $\Delta_{1,2}(x)$, estimated by assuming BCS superconductors with the given transition temperatures and find the solutions for $\theta_{1,2}(x,\omega_n)$. 
Using these solutions, the new pair potentials $\Delta_{1,2}(x)$ are found from Eq.~\ref{eqn:dlog}. 
The iterations are repeated until convergence in $\Delta_{1,2}(x)$ ($\delta \Delta_{1,2} / \Delta_{1,2} < 0.001\%$) is achieved. 
Next, after the pair potentials are determined, we solve Eq.~\ref{eqn:usadel} on the energy axis, $E$, by replacing $\hbar\omega_n$ with $-jE$. 
This method provides $\theta_{1,2}(x,E)$ for calculating spatial- and energy-resolved densities of states in both layers. 
Subsequently, we calculate the local values of the complex conductivity with Eqs.~\ref{eqn:sigma1} and~\ref{eqn:sigma2}, and provide values for the complex impedance $Z_S$, quality factor $Q_S$, and surface inductance $L_S$. 
\begin{center}
\begin{table}[htbp]
  \begin{threeparttable}[b]
    \caption{Physical Parameters for Calculations}
    \label{tab:table9}
    \centering
    \begin{tabular}{| p{0.20\textwidth} | p{0.06\textwidth} | p{0.06\textwidth} | p{0.06\textwidth} | }
     \hline 
     Parameter & Al & Ir & Cu\\
     \hline
       Transition temperature $T_C$ & 1.52 K \tnote{a} & 0.27 K \tnote{a} & 1.0 pK \tnote{b} \\
       Electron phonon coupling constant $\lambda$ \tnote{c, d, e, f}  & 0.44 & 0.34 & 0.08 \\
       Electronic specific heat coefficient $\gamma$ in JK$^{-2}$m$^{-3}$ \tnote{e, g} & $135.0 $ & 370.0 & 98.0 \\
       Fermi wavelength $\lambda_f$ in nm \tnote{h, i}  & 0.360 & 0.536 & 0.455 \\
       Resistivity $\rho$ in $\mu \Omega$cm at low temperatures \tnote{a}  & 3.15 & 13.2 & 2.15 \\
       Debye temperature $T_D$ in Kelvin \tnote{c, e, j, k} & 428 & 420 & 335 \\
       Coherence length $\xi$ in nm \tnote{l} & 384.2 & 90.3 & 451.2 \\
      \hline
    \end{tabular}
    \begin{tablenotes}
      \item [a] Averaged values of 20\textendash30 nm thin films fabricated and measured at Argonne; \item [b] Assumed small value. See Fig.~\ref{fig:2} for comparison of results at three Cu transition temperatures; \item [c]~\cite{Lin:08}; \item [d]~\cite{Allen:87}; \item [e]~\cite{McMillan:68}; \item [f]~\cite{Brorson:90}; \item [g]~\cite{Irwin:05}; \item [h]~\cite{Reale:74}; \item [i]~\cite{Lin:15}; \item [j]~\cite{Kittel:96}; \item [k]~\cite{Tong:19};  \item [l] $\xi = \sqrt( \hbar D /2 \pi k_B T)$, $T=100$~mK. 
    \end{tablenotes}
  \end{threeparttable}
\end{table}
\end{center}

\begin{figure}[tbp]
\centering
\includegraphics[width=3.0in]{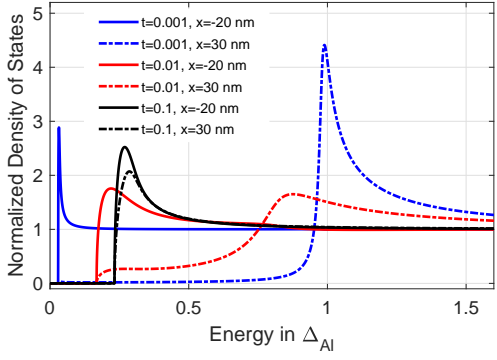}
\caption{Densities of states of a bilayer made of 20 nm Cu ($x<0$) and 30 nm Al ($x>0$) at three electron transmission coefficients. $T=100$~mK.}
\label{fig:1}
\end{figure}

We limit our calculations to thin film Ir/Al and Cu/Al bilayers on high resistivity silicon wafers deposited using traditional sputter deposition. 
The physical input parameters are summarized in Table~\ref{tab:table9}. 
The only free parameter is the electron transmission coefficient $t$ at the interface of a bilayer. 
It is related to the specific interface resistance, which is given by $R_B=A/G_{int}=\lambda_f^2 / 8tG_K$. 
A small electron transmission coefficient corresponds to a large specific interface resistance. 
Fig.\ref{fig:1} shows the density of states of a thin Cu/Al bilayer at three electron transmission coefficients. 
When $t=0.001$ or $0.01$, the specific interface resistance is large and electrons have little chance to pass through the interface, therefore, the proximity effect between Cu and Al is weak. 
In Cu, the density of states peaks at a low energy. 
In Al, the density of states peaks near $\Delta_{Al}(T=0) \approx 231~\mu$eV and there is a mini gap aligned with the lower edge of the density of states in Cu. 
When $t=0.1$, which corresponds to a metallic contact estimated from experimental $T_C$ data of proximity bilayers~\cite{Martinis:00, Hennings-Yeomans:20}, the distributions of densities of states in Cu and Al merge together with only one lower edge. 
We assume $t=0.1$ in the rest of the paper.
\begin{figure}[htbp]
\centering
\includegraphics[width=3.0in]{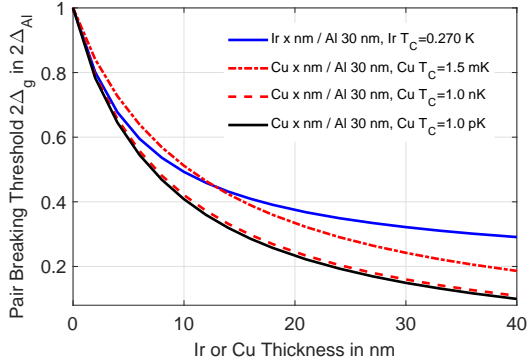}
\caption{Pair breaking threshold of a 30 nm Al-based bilayer as a function of the thickness of a proximity Ir or Cu layer. $t=0.1$. $T=100$~mK.}
\label{fig:2}
\end{figure}

We investigate the pair breaking threshold tuning of an Al film proximitized with an Ir or Cu film as shown in Fig.~\ref{fig:2}. 
The $T_C$ of Ir is listed in Table~\ref{tab:table9}. 
Though there is no measured $T_C$ for Cu yet, we cannot use $T_C=0$~K for the calculations with Eq.~\ref{eqn:dlog}. 
$T_C \approx 1.5$~mK is extracted based on a BCS superconductor~\cite{Tinkham:94} with an electron phonon coupling constant $\lambda=0.08$~\cite{Brorson:90} by using $T_C=1.134 T_D e^{-1/\lambda}$, where $T_D$ is the Debye temperature of Cu. 
$T_C=1.0$~nK and $T_C=1.0$~pK are also used for comparison. 
When $T_C<1.0$~nK, the calculation should have little difference if the $T_C$ of Cu is vanishingly small. 
Fig.~\ref{fig:2} demonstrates that the pair breaking threshold ($2\Delta_g$) of an Al film can be effectively tuned downward with the deposition of an Ir or Cu film. 
The dependence of $2\Delta_g$ of a Cu/Al bilayer on Cu thickness is very similar to that of $T_C$ of an Ir/Pt bilayer on Pt thickness~\cite{Wang:17}. 
It can approach zero at a large Cu film thickness. 
However, the lower bound of $2\Delta_g$ is limited to above the energy gap of the Ir film for an Ir/Al bilayer. 
\begin{figure}[tbp]
\centering
\includegraphics[width=3.0in]{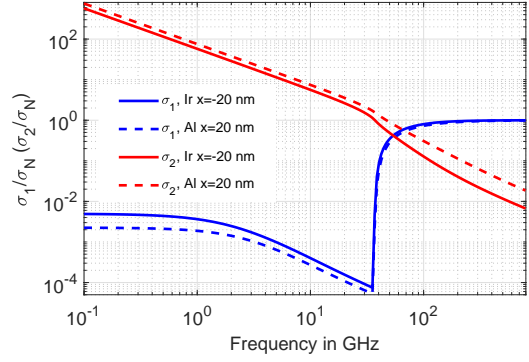}
\caption{Complex conductivities at the outer edges of a bilayer made of 20 nm Ir ($x<0$) and 20 nm Al ($x>0$). $t=0.1$. $T=100$~mK.
}
\label{fig:3}
\end{figure}
\begin{figure}[tbp]
\centering
\includegraphics[width=3.0in]{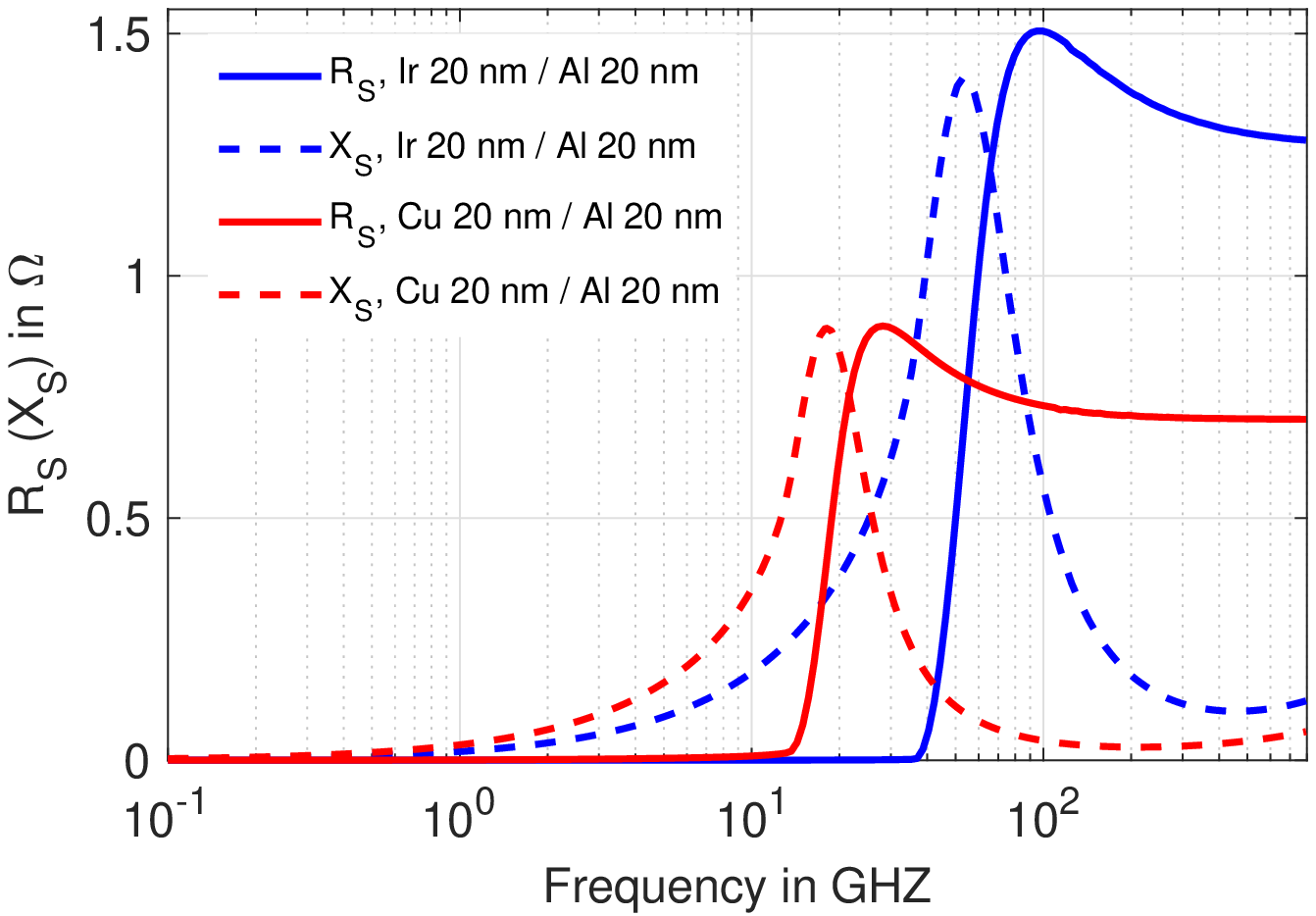}
\includegraphics[width=3.0in]{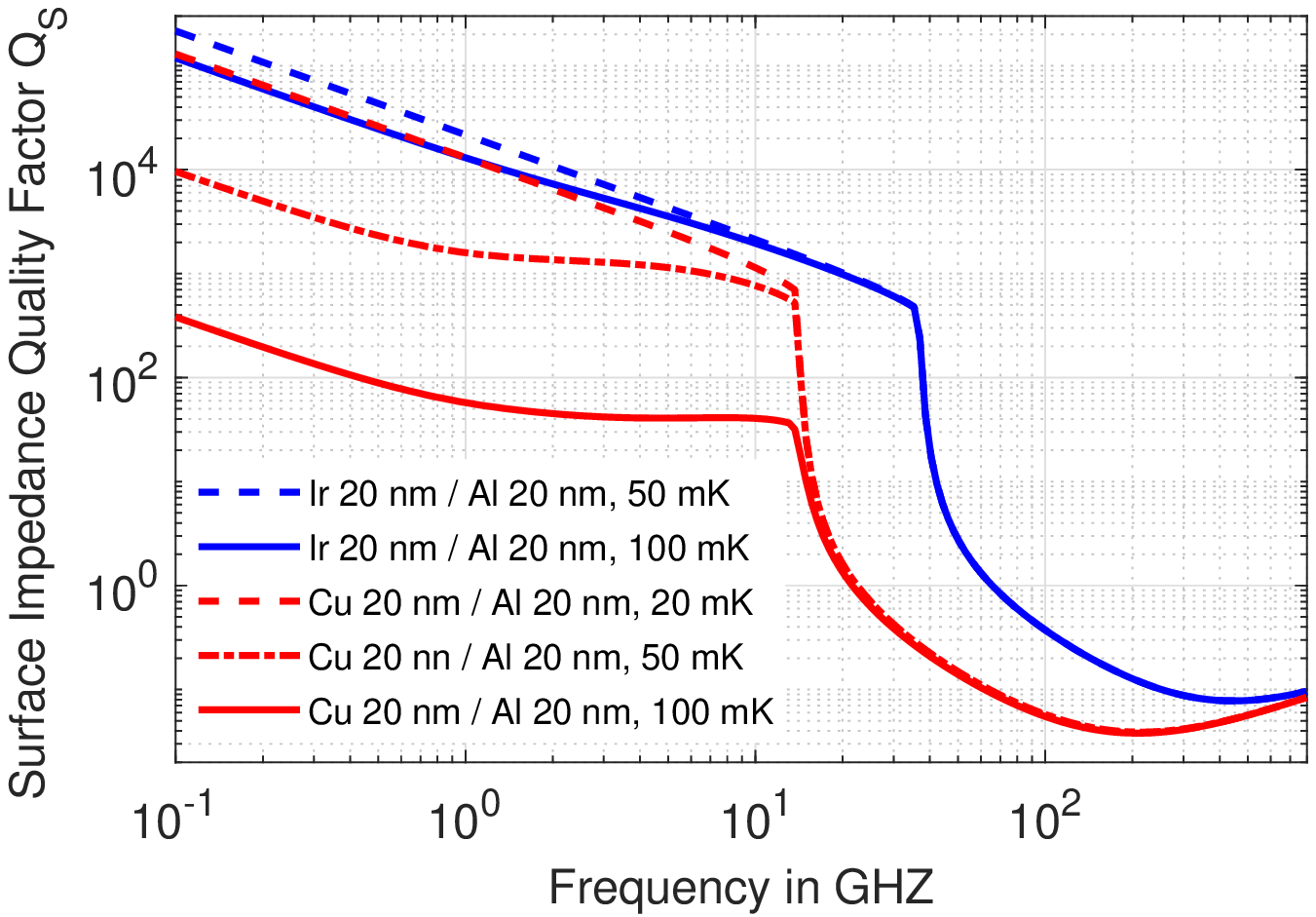}
\caption{Complex impedances of two thin bilayers (top). $t=0.1$. $T=100$~mK. 
The frequency dependence of surface impedance quality factors of the two thin bilayers at several temperatures (bottom). $t=0.1$.}
\label{fig:4}
\end{figure}

Fig.~\ref{fig:3} shows the real and imaginary parts of complex conductivities at the outer edges of a 20 nm Ir and 20 nm Al bilayer. 
In both layers, the small real part $\sigma_1 / \sigma_N$ decreases as a function of electromagnetic wave frequency until the pair breaking threshold $f_g = 2 \Delta_g / h$ is approached. 
Then it rises sharply at $f_g$ and rapidly reaches the normal state conductivity. 
The frequency, $f_g$, is the pair breaking frequency above which radiation is energetic enough to break Cooper pairs. 
The imaginary part $\sigma_2 / \sigma_N$ also decreases as a function of frequency, however, it decreases even faster above $f_g$. 
Overall, $\sigma_2 / \sigma_N \gg \sigma_1 / \sigma_N$ when $f < f_g$.

With complex conductivities, the surface impedance of a bilayer can be calculated with Eqs.~\ref{eqn:matrix} and~\ref{eqn:Zs}. 
The upper panel of Fig.~\ref{fig:4} shows the real and imaginary parts of the surface impedance for two bilayer configurations: 1) 20 nm Ir and 20 nm Al, and 2) 20 nm Cu and 20 nm Al. 
Above the pair breaking threshold $f_g$, the Ir/Al bilayer has a larger surface impedance relative to the Cu/Al bilayer because the resistivity of Ir is larger than that of Cu. 
Below the pair breaking threshold $f_g$, the imaginary part depends exponentially on frequency and it is much larger than the dissipative part, which decreases exponentially with decreasing frequency and is very close to zero. 
In general, the frequency dependence of the surface impedance of the bilayers is similar to that of the Al/Ti/Al trilayers~\cite{Zhao:17}. 

The ratio of the imaginary part to the dissipative part of surface impedance, which is defined as surface impedance quality factor $Q_S$ in Eq.~\ref{eqn:Qs}, is a useful parameter to understand the resistive loss of a resonator~\cite{Zmuidzinas:12}. 
In the lower panel of Fig.~\ref{fig:4}, the blue dashed line and solid line are for the Ir/Al bilayer, which has a pair breaking threshold of about 35.6 GHz. 
The red dashed line, dash-dotted line and solid line are for the thin Cu/Al bilayer, which has a pair breaking threshold of about 13.7 GHz. 
The surface impedance quality factor strongly depends on pair breaking threshold, excitation electromagnetic wave frequency, and operational temperature. 
When the pair breaking threshold is small, the surface impedance quality factor is small at 100~mK. 
To have a larger quality factor, a resonator with such as a small pair breaking threshold needs to be operated at a lower frequency and a lower temperature.
\begin{figure}[htbp]
\centering
\includegraphics[width=3.0in]{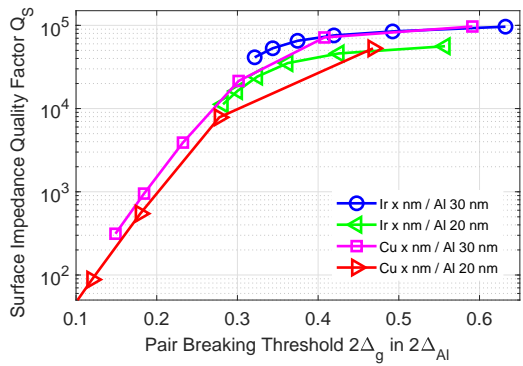}
\caption{Surface impedance quality factor vs pair breaking threshold of Ir/Al and Cu/Al bilayers at 0.5 GHz. Starting from the largest $2\Delta_g$, the thickness of Ir or Cu increases from 5 to 30 nm with a step of 5 nm. $t=0.1$. $T=100$~mK.}
\label{fig:5}
\end{figure}
\begin{figure}[htbp]
\centering
\includegraphics[width=3.0in]{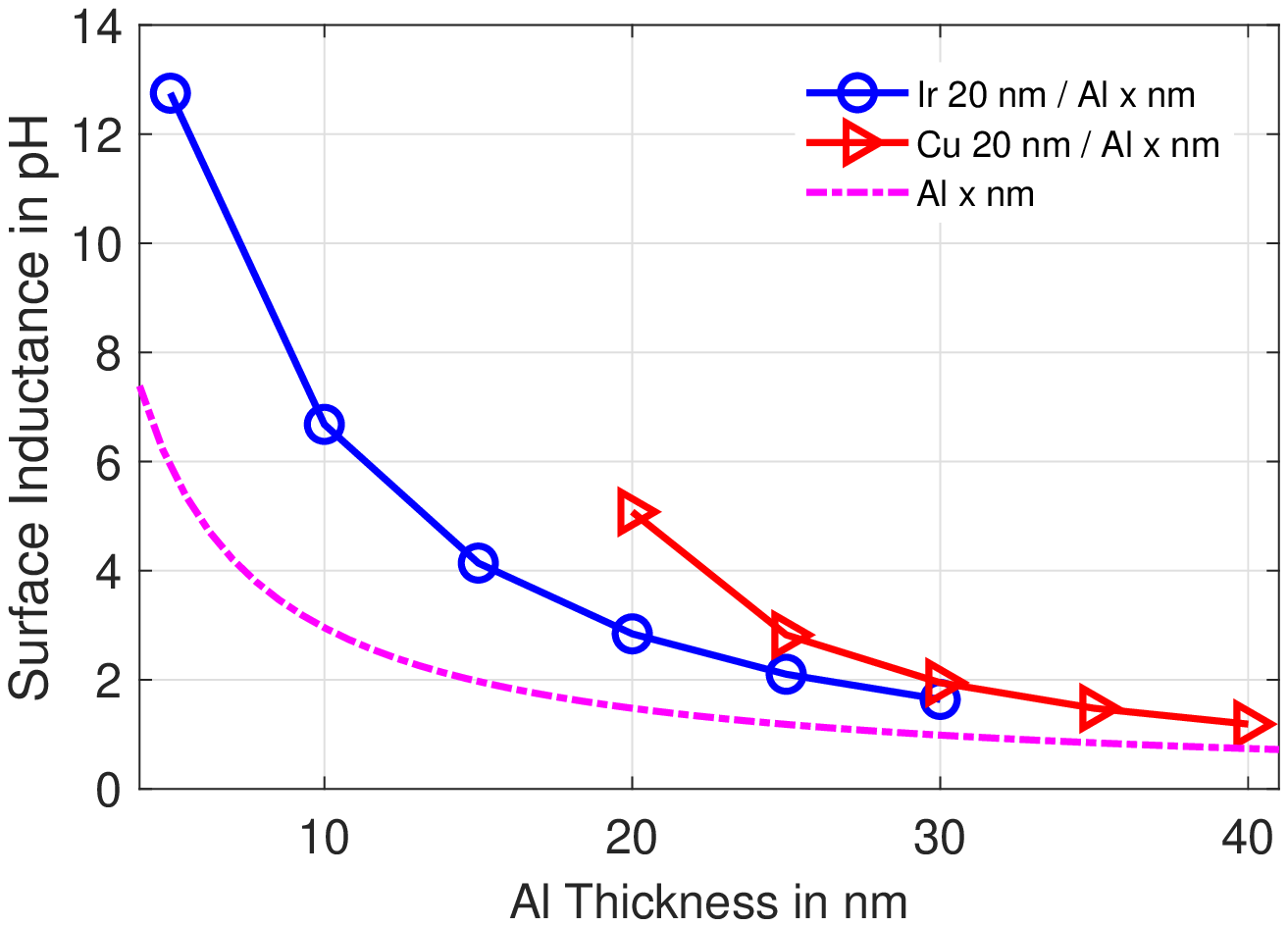}
\includegraphics[width=3.0in]{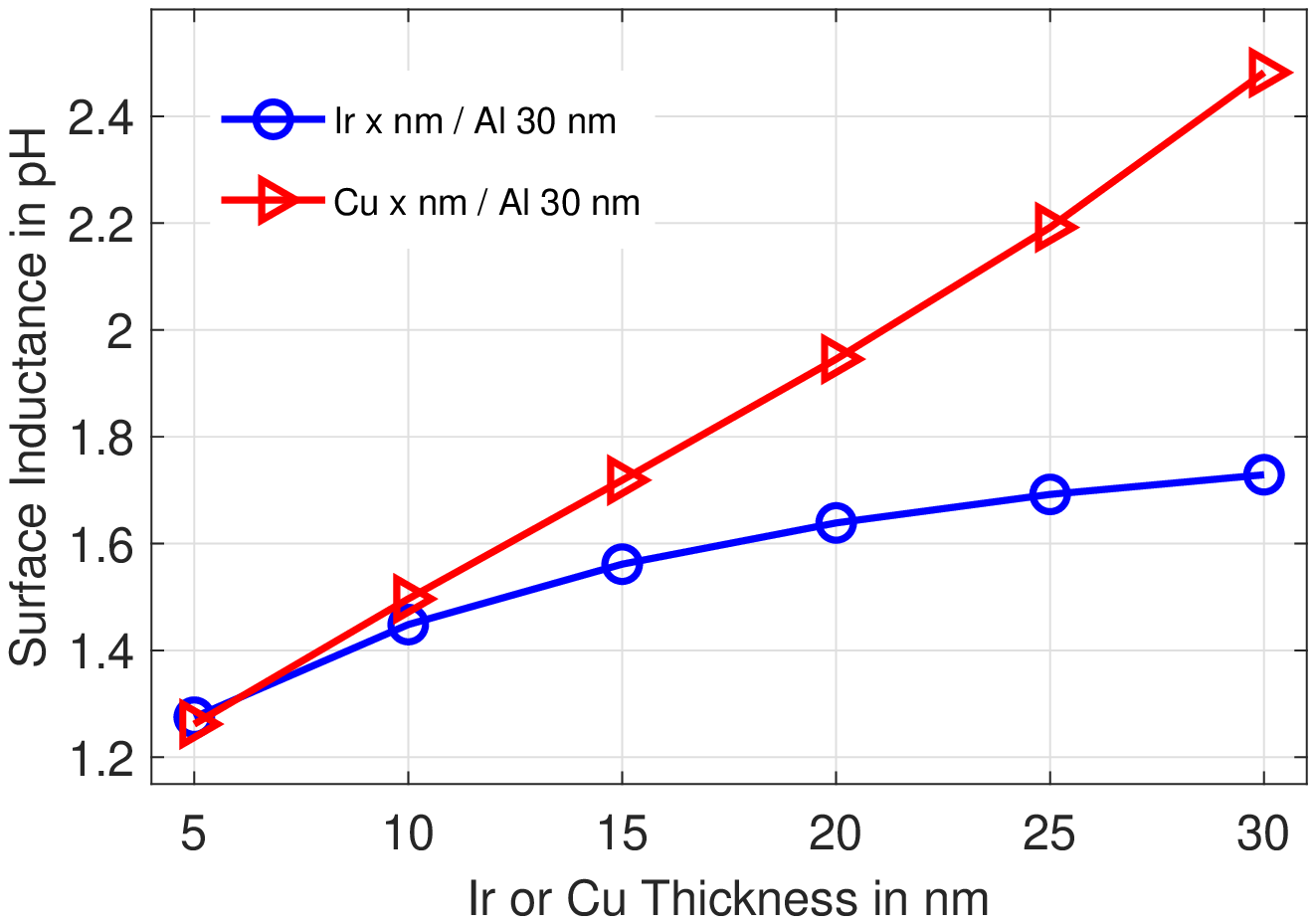}
\caption{Surface inductance dependence on Al film thickness (top) and on Ir or Cu film thickness (bottom). $t=0.1$. $T=100$~mK.}
\label{fig:6}
\end{figure}

Four calculated data sets of surface impedance quality factor $Q_S$ of Ir/Al and Cu/Al bilayers at 0.5 GHz and at 100 mK are summarized in Fig.~\ref{fig:5}. There are several conclusions we can draw. 
First, when the pair breaking threshold $2\Delta_g$ is small, the $Q_S$ of an Al-based bilayer depends on $2\Delta_g$ exponentially. 
This dependence is similar to that of $Q_S$ on $2\Delta_g$ for a single layer superconductor film~\cite{Zmuidzinas:12}. 
The decrease of calculated $Q_S$ with an increasing thickness of an Ir or Cu proximity film qualitatively agrees the measured surface resistance of Nb/Cu bilayers~\cite{Pambianchi:96}, which increases many orders of magnitude with Cu layer increasing from ten to a few tens of nm. 
Next, for bilayers with the same Al film thickness regardless of whether they are Ir/Al or Cu/Al, the surface impedance quality factor decreases monotonically with decreasing $2\Delta_g$. 
In addition, at the same $2\Delta_g$, a bilayer with 30 nm Al film has a larger $Q_S$ than a bilayer with 20 nm Al film does.

The upper panel of Fig.~\ref{fig:6} shows the dependence of surface inductance $L_S$, estimated at 0.5 GHz, of a Cu/Al bilayer and an Ir/Al bilayer on Al film thickness for a given Ir or Cu film thickness of 20 nm. The dash-dotted line is for a single Al film estimated with Eq.~6 in Ref.~\cite{Annunziata:10} for comparison. The $L_S$ of an Al-based bilayer is enhanced with additional Ir or Cu proximity layer. These results are consistent with the measured magnetic penetration depth enhancement of an Nb film proximitized with an Al or Cu film~\cite{Pambianchi:95, Pambianchi:96, Wangr:00}. 
In addition, the $L_S$ of the bilayers is inversely proportional to the thickness of Al film. 
The lower panel of Fig.~\ref{fig:6} shows a monotonic increase of $L_S$ with thickness of an Ir or Cu film at a given Al film thickness of 30 nm. This is consistent with the measured dependence of magnetic penetration depth of Nb-based bilayers on the thickness of an Al or Cu film~\cite{Pambianchi:95, Pambianchi:96, Wangr:00}. However, the $L_S$ of a Cu/Al bilayer is much larger than that of an Ir/Al bilayer at the same large thickness of Ir and Cu films. This could indicate that the $L_S$ is strongly correlated to the $2\Delta_g$, whose dependence on the thickness of an Ir or Cu film is shown in Fig~\ref{fig:2}.

\section{Conclusion}
By numerically solving the Usadel equations, we have calculated the pair breaking threshold, $2\Delta_g$, of Ir/Al and Cu/Al thin-film bilayers. 
Across a thin bilayer in metallic contact, there is only a single $2\Delta_g$ defined by the density of states. 
The value of $2\Delta_g$ can be monotonically tuned by changing the relative thicknesses of Al and Ir (or Cu) films. 
Utilizing the Nam's expressions for complex conductivity, we have studied the electromagnetic properties of these bilayers. 
The surface inductance, $L_S$, of an Al-based bilayer is larger than that of the bare Al film only. 
To further enhance $L_S$, a bilayer should have a thin Al film and/or a thick Ir (or Cu) film. 
In general, the surface impedance quality factor $Q_S$ of an Al-based bilayer depends on $2\Delta_g$ in the similar way as the $Q_S$ of a single layer Al film does. 
However, for two bilayers with the same $2\Delta_g$, the bilayer with thicker Al film is expected to have a larger $Q_S$. 
Moreover, the $Q_S$ of an Ir/Al bilayer can be larger than that of a Cu/Al bilayer with the same thicknesses because the Ir/Al has a larger $2\Delta_g$. 
In the case of a small $2 \Delta_g$ for a low threshold (10-20 GHz) detection, a KID made of an Al-based bilayer needs to be operated at a low temperature (10 mK) and with a low readout frequency ($<$2.0 GHz) for a large quality factor. 
The results of this work can be used to aid in the design and experiment of Al-based bilayer KIDs.


\begin{thebibliography}{1}
\bibliographystyle{IEEEtran}

\bibitem{Zmuidzinas:12}
J. Zmuidzinas, ``Superconducting microresonators: Physics and applications,'' \textit{Annu. Rev. Condens. Matter Phys}, vol. 3, no. 1 (2012). DOI: 10.1146/annurev-conmatphys-020911-125022.

\bibitem{Austermann:18}
J. E. Austermann, J. A. Beall, S. A. Bryan, B. Dober, J. Gao, G. Hilton, J. Hubmayr et al., ``Millimeter-wave polarimeters using kinetic inductance detectors for TolTEC and beyond,'' \textit{Journal of Low Temperature Physics}, vol. 193, no. 3 (2012). DOI: 10.1007/s10909-018-1949-5. 

\bibitem{Dober:14}
B. J. Dober, P. A. R. Ade, P. Ashton, F. E. Angil\`{e}, J. A. Beall, D. Becker, K. J. Bradford et al., ``The next-generation BLASTPol experiment,'' \textit{ Millimeter, Submillimeter, and Far-Infrared Detectors and Instrumentation for Astronomy VII}, vol. 9153, pp.137-148, SPIE, 2014. DOI: 10.1117/12.2054419. 

\bibitem{Karkare:22}
K. S. Karkare, A. J. Anderson, P. S. Barry, B. A. Benson, J. E. Carlstrom, T. Cecil, C. L. Chang et al., ``SPT-SLIM: A Line Intensity Mapping Pathfinder for the South Pole Telescope,'' \textit{ Journal of Low Temperature Physics}, (2022): 1-8. DOI: 10.1007/s10909-022-02702-2. 

\bibitem{Meeker:18}
S. R. Meeker, B. A. Mazin, A. B. Walter, P. Strader, N. Fruitwala, C. Bockstiegel, P. Szypryt et al., ``DARKNESS: a microwave kinetic inductance detector integral field spectrograph for high-contrast astronomy,'' \textit{Publications of the Astronomical Society of the Pacific}, vol. 130, no. 988 (2018): 065001. DOI: 10.1088/1538-3873/aab5e7. 
 
\bibitem{Walter:20}
A. B. Walter, N. Fruitwala, S. Steiger, J. I. Bailey, N. Zobrist, N. Swimmer, I. Lipartito et al., ``The MKID exoplanet camera for Subaru SCExAO,'' \textit{Publications of the Astronomical Society of the Pacific}, vol. 132, no. 1018 (2020): 125005. DOI: 10.1088/1538-3873/abc60f. 
 
\bibitem{Calvo:16}
M. Calvo, A. Beno\^{i}t, A. Catalano, J. Goupy, A. Monfardini, N. Ponthieu, E. Barria et al., ``The NIKA2 instrument, a dual-band kilopixel KID array for millimetric astronomy,'' \textit{Journal of Low Temperature Physics}, vol. 184, no. 3 (2016): 816-823. DOI: 10.1007/s10909-016-1582-0. 

\bibitem{Moore:12}
D. C. Moore, S. R. Golwala, B. Bumble, B. Cornell, P. K. Day, H. G. LeDuc, and J. Zmuidzinas, ``Position and energy-resolved particle detection using phonon-mediated microwave kinetic inductance detectors,'' \textit{Applied Physics Letters}, vol. 100, no. 23 (2012): 232601. DOI: 10.1063/1.4726279. 

\bibitem{Cornell:14}
B. Cornell, D. C. Moore, S. R. Golwala, B. Bumble, P. K. Day, H. G. LeDuc, and J. Zmuidzinas, ``Particle Detection Using MKID-Based Athermal-Phonon Mediated Detectors,'' \textit{Journal of Low Temperature Physics}, vol. 176, no. 5 (2012): 891-897. DOI: 10.1007/s10909-013-1039-7. 
 
\bibitem{Cardani:18}
L. Cardani, N. Casali, A. Cruciani, H. Le Sueur, M. Martinez, F. Bellini, M. Calvo et al., ``Al/Ti/Al phonon-mediated KIDs for UV–vis light detection over large areas,'' \textit{Superconductor Science and Technology}, vol. 31, no. 7 (2018): 075002. DOI: 10.1088/1361-6668/aac1d4. 
 
\bibitem{Colantoni:20}
I. Colantoni, C. Bellenghi, M. Calvo, R. Camattari, L. Cardani, N. Casali, A. Cruciani et al., ``Bullkid: Bulky and low-threshold kinetic inductance detectors,'' \textit{Journal of Low Temperature Physics}, vol. 199, no. 3 (2020): 593-597. DOI: 10.1007/s10909-020-02408-3. 

\bibitem{Miceli:14}
A. Miceli, T. W. Cecil, L. Gades, and O. Quaranta., ``Towards x-ray thermal kinetic inductance detectors,'' \textit{Journal of Low Temperature Physics}, vol. 176, no. 3 (2014): 479-503. DOI: 10.1007/s10909-013-1033-0.

\bibitem{Ulbricht:15}
G. Ulbricht, B. A. Mazin, P. Szypryt, A. B. Walter, C. Bockstiegel, and B. Bumble, ``Highly multiplexible thermal kinetic inductance detectors for x-ray imaging spectroscopy,'' \textit{Applied Physics Letters}, vol. 106, no. 25 (2015): 479-503. DOI: 10.1063/1.4923096.
 
\bibitem{Wandui:20}
A. Wandui, J. J. Bock, C. Frez, M. Hollister, L. Minutolo, H. Nguyen, B. Steinbach, A. Turner, J. Zmuidzinas, and R. O'Brient, ``Thermal kinetic inductance detectors for millimeter-wave detection,'' \textit{Journal of Applied Physics}, vol. 128, no. 4 (2020): 044508. DOI: 10.1063/5.0002413.

\bibitem{Vissers:15}
M. R. Vissers, J. Hubmayr, M. Sandberg, S. Chaudhuri, C. Bockstiegel, and J. Gao, ``Frequency-tunable superconducting resonators via nonlinear kinetic inductance,'' \textit{Applied Physics Letters}, vol. 107, no. 6 (2015): 062601. DOI: 10.1063/1.4927444.

\bibitem{Kher:217}
A. S. Kher, ``Superconducting nonlinear kinetic inductance devices,'' PhD disssertation, California Institute of Technology, 2017.

\bibitem{Luomahaara:14}
J. Luomahaara, V. Vesterinen, L. Gr\"{o}nberg, and J. Hassel, ``Kinetic inductance magnetometer,'' \textit{Nature communications}, vol. 5, no. 1 (2014): 1-7. DOI: 10.1038/ncomms5872.

\bibitem{Vissers:13}
M. R. Vissers, J. Gao, M. Sandberg, S. M. Duff, D. S. Wisbey, K. D. Irwin, and D. P. Pappas, ``Proximity-coupled Ti/TiN multilayers for use in kinetic inductance detectors,'' \textit{Applied Physics Letters}, vol. 102, no. 23 (2013): 232603. DOI: 10.1063/1.4804286. 

\bibitem{Hu:20}
J. Hu, M. Salatino, A. Traini, C. Chaumont, F. Boussaha, C. Goupil, and M. Piat, ``Proximity-coupled Al/Au bilayer kinetic inductance detectors,'' \textit{Journal of Low Temperature Physics}, vol. 199, no. 1 (2020): 355-361. DOI: 10.1007/s10909-019-02313-4. 

\bibitem{Catalano:15}
A. Catalano, J. Goupy, H. L. Sueur, A. Benoit, O. Bourrion, M. Calvo, A. D’addabbo et al., ``Bi-layer kinetic inductance detectors for space observations between 80–120 GHz,'' \textit{Astronomy \& Astrophysics}, 580 (2015): A15. DOI: 10.1051/0004-6361/201526206. 

\bibitem{Aja:20}
B. Aja, L. De La Fuente, A. Fernandez, J. P. Pascual, E. Artal, M. C. De Ory, M. T. Magaz, D. Granados, J. Martin-Pintado, and A. Gomez, ``Bi-layer kinetic inductance detectors for W-band,'' \textit{ IEEE/MTT-S International Microwave Symposium (IMS)}, pp. 932-935. IEEE, 2020. DOI: 10.1109/IMS30576.2020.9223828.

\bibitem{Sekine:14}
M. Sekine, Y. Sekimoto, T. Noguchi, A. Miyachi, K. Karatsu, T. Nitta, S. Sekiguchi, T. Okada, and M. Naruse, ``Development of Nb/Al bi-layer MKID camera--Temperature dependence of quality factor,'' \textit{IEICE Technical Report; IEICE Tech. Rep.}, vol. 133, no. 401 (2014): 73-77. 

\bibitem{Zobrist:22}
N. Zobrist, W. H. Clay, G. Coiffard, M. Daal, N. Swimmer, P. Day, and B. A. Mazin, ``Membraneless phonon trapping and resolution enhancement in optical microwave kinetic inductance detectors,'' \textit{Physical Review Letters}, vol. 129, no. 1 (2022): 017701. DOI: 10.1103/PhysRevLett.129.017701.
 
\bibitem{Claassen:99}
J. H. Claassen, S. Adrian, and R. J. Soulen, ``Large non-linear kinetic inductance in superconductor/normal metal bilayer films,'' \textit{IEEE transactions on applied superconductivity}, vol. 9, no. 2 (1999): 4189-4192. DOI: 10.1109/77.783948.

\bibitem{Ustavschikov:20}
S. S. Ustavschikov, M. Yu Levichev, I. Yu Pashenkin, A. M. Klushin, and D. Yu Vodolazov, ``Approaching depairing current in dirty thin superconducting strip covered by low resistive normal metal,'' \textit{Superconductor Science and Technology}, vol. 34, no. 1 (2020): 015004. DOI: 10.1088/1361-6668/abc2ad.

\bibitem{Mashian:15}
N. Mashian, A. Sternberg, and A. Loeb, ``Predicting the intensity mapping signal for multi-J CO lines,'' \textit{Journal of Cosmology and Astroparticle Physics}, 2015, no. 11 (1995): 028. DOI: doi: 10.1088/1475-7516/2015/11/028.
 
\bibitem{Dizgah:22}
A. M. Dizgah, G. K. Keating, K. S. Karkare, A. Crites, and S. R. Choudhury, ``Neutrino Properties with Ground-Based Millimeter-Wavelength Line Intensity Mapping,'' \textit{The Astrophysical Journal}, vol. 926, no. 2 (2022): 1137. DOI: 10.3847/1538-4357/ac3edd.

\bibitem{Usadel:70}
K. D. Usadel, ``Generalized diffusion equation for superconducting alloys,'' \textit{Physical Review Letters}, vol. 25, no. 8 (1970): 507. DOI: 10.1103/PhysRevLett.25.507.
 
\bibitem{Golubov:04}
A. A. Golubov, M. Yu Kupriyanov, and E. Il’Ichev, ``The current-phase relation in Josephson junctions,'' \textit{Reviews of modern physics}, vol. 76, no. 2 (2004): 411. DOI: 10.1103/RevModPhys.76.411.
         
\bibitem{Golubov:95}
A. A. Golubov, E. P. Houwman, J. G. Gijsbertsen, V. M. Krasnov, J. Flokstra, H. Rogalla, and M. Yu Kupriyanov, ``Proximity effect in superconductor-insulator-superconductor Josephson tunnel junctions: Theory and experiment,'' \textit{Physical Review B}, vol. 51, no. 2 (1995): 1073. DOI: 10.1103/physrevb.51.1073.

\bibitem{Zhao:99}
S. P. Zhao and Q. S. Yang, ``Penetration depth in conventional layered superconductors: A proximity-effect model,'' \textit{Physical Review B}, vol. 59 (22), (1999): 14630. DOI: 10.1103/PhysRevB.59.14630.

\bibitem{Brammertz:01}
G. Brammertz, A. Poelaert, A. A. Golubov, P. Verhoeve, A. Peacock, and H. Rogalla, ``Generalized proximity effect model in superconducting bi-and trilayer films,'' \textit{Journal of Applied Physics}, vol. 90, no. 1 (2001): 355-364. DOI: 10.1063/1.1376411.

\bibitem{Zhao:17}
S. Zhao, D. J. Goldie, S. Withington, and C. N. Thomas, ``Exploring the performance of thin-film superconducting multilayers as kinetic inductance detectors for low-frequency detection,'' \textit{Superconductor Science and Technology}, vol. 31, no. 1 (2017): 015007. DOI: 10.1088/1361-6668/aa94b7.

\bibitem{John:84}
W. John, V. V. Nemoshkalenko, V. N. Antonov, and Vl. N. Antonov, ``Electron-Phonon Interaction in Transition Metals. Results of Relativistic APW Band Structure Calculations,'' \textit{physica status solidi (b)}, vol. 121, no. 1 (1984): 233-239. DOI: 10.1002/pssb.2221210125.
 
\bibitem{Kuprianov:88}
M. Yu Kuprianov, and V. F. Lukichev, ``Influence of boundary transparency on the critical current of dirty SS'S structures,'' \textit{Zh. Eksp. Teor. Fiz}, vol. 94, (1988): 149. 

\bibitem{Fominov:01}
Ya V. Fominov and M. V. Feigel'man, ``Superconductive properties of thin dirty superconductor–normal-metal bilayers,'' \textit{Physical Review B}, vol. 63, no. 9 (2001): 094518. DOI: 10.1103/PhysRevB.63.094518.

\bibitem{Zhu:09}
S. Zhu, T. Zijlstra, A. A. Golubov, M. Van den Bemt, A. M. Baryshev, and T. M. Klapwijk, ``Magnetic field dependence of the coupling efficiency of a superconducting transmission line due to the proximity effect,'' \textit{Applied physics letters}, vol. 95, no. 25 (2001): 253502. DOI: 10.1063/1.3276076. 

\bibitem{Martinis:00}
J. M. Martinis, G. C. Hilton, K. D. Irwin, and D. A. Wollman, ``Calculation of Tc in a normal-superconductor bilayer using the microscopic-based Usadel theory,'' \textit{Nuclear Instruments and Methods in Physics Research Section A: Accelerators, Spectrometers, Detectors and Associated Equipment}, vol. 444, no. 1-2 (2000): 23-27. DOI:  /10.1016/S0168-9002(99)01320-0.
                
\bibitem{Nam:67}
S. B. Nam, ``Theory of electromagnetic properties of strong-coupling and impure superconductors II,'' \textit{ Physical Review}, Vol. 156, no. 2 (1967): 487. DOI: 10.1103/PhysRev.156.487.


\bibitem{Lin:08}
Z. Lin, L. V. Zhigilei, and V. Celli, ``Electron-phonon coupling and electron heat capacity of metals under conditions of strong electron-phonon nonequilibrium,'' \textit{Physical Review B}, vol. 77 no. 7 (2008): 075133. DOI: 10.1103/PhysRevB.77.075133.

\bibitem{Allen:87}
P. B. Allen, ``Empirical electron-phonon $\lambda$ values from resistivity of cubic metallic elements,'' \textit{Physical Review B}, vol. 36, no. 5 (1987): 2920. DOI: 10.1103/PhysRevB.36.2920.

\bibitem{McMillan:68}
W. L. McMillan, ``Transition temperature of strong-coupled superconductors,'' \textit{Physical Review}, vol. 167, no. 2 (1968): 331. DOI: 10.1103/PhysRev.167.331. 

\bibitem{Brorson:90}
S. D. Brorson, A. Kazeroonian, J. S. Moodera, D. W. Face, T. K. Cheng, E. P. Ippen, M. S. Dresselhaus, and G. Dresselhaus, ``Femtosecond room-temperature measurement of the electron-phonon coupling constant $\gamma$ in metallic superconductors,'' \textit{Physical Review Letters}, vol. 64, no. 18 (1990): 2172. DOI: 10.1103/PhysRevLett.64.2172.

\bibitem{Irwin:05}
K. D. Irwin and G. C. Hilton, {\it{Transition-edge sensors}} in Cryogenic particle detection, pp. 63-150. Springer, Berlin, Heidelberg, 2005. 

\bibitem{Reale:74}
C. Reale, ``Conductivity data for the transition metals derived from considerations on the charge transport in thin films,'' \textit{Journal of Physics F: Metal Physics}, vol. 4, no. 12 (1974): 2218. DOI: 10.1088/0305-4608/4/12/017.

\bibitem{Lin:15}
S. W. Lin, Y. H. Wu, L. Chang, C. T. Liang, and S. D. Lin, ``Pure electron-electron dephasing in percolative aluminum ultrathin film grown by molecular beam epitaxy,'' \textit{Nanoscale Research Letters}, vol. 10, no. 1 (2015): 1-7. DOI 10.1186/s11671-015-0782-x.

\bibitem{Kittel:96}
C. Kittel, {\it{Introduction to Solid State Physics}}. 7th Ed., Hoboken, NJ., USA. Wiley, 1995.

\bibitem{Tong:19}
Z. Tong, S. Li, X. Ruan, and H. Bao, ``Comprehensive first-principles analysis of phonon thermal conductivity and electron-phonon coupling in different metals,'' \textit{Physical Review B}, vol. 100, no. 14 (2019): 144306. DOI: 10.1103/PhysRevB.100.144306.


\bibitem{Hennings-Yeomans:20}
R. Hennings-Yeomans, C. L. Chang, J. Ding, A. Drobizhev, B. K. Fujikawa, S. Han, G. Karapetrov et al., ``Controlling Tc of iridium films using the proximity effect,'' \textit{Journal of Applied Physics}, vol. 128, no. 15 (2020): 154501. DOI:  10.1063/5.0018564.

\bibitem{Tinkham:94}
M. Tinkham, {\it{Introduction to Superconductivity}}. McGraw-Hill, New York, 2nd edition,1994.

\bibitem{Wang:17}
G. Wang, J. Beeman, C. L. Chang, J. Ding, A. Drobizhev, B. K. Fujikawa, K. Han et al., ``Modeling iridium-based trilayer and bilayer transition-edge sensors,'' \textit{IEEE Transactions on Applied Superconductivity}, vol. 27, no. 4 (2017): 1-5. DOI: 10.1109/TASC.2016.2646373.

\bibitem{Annunziata:10}
A. J. Annunziata, D. F. Santavicca, L. Frunzio, G. Catelani, M. J. Rooks, A. Frydman, and D. E. Prober, ``Tunable superconducting nanoinductors,'' \textit{Nanotechnology}, vol. 21, no. 44 (2010): 445202. DOI: 10.1088/0957-4484/21/44/445202.

\bibitem{Pambianchi:95}
M. S. Pambianchi, S. N. Mao, and S. M. Anlage, ``Microwave surface impedance of proximity-coupled Nb/Al bilayer films,'' \textit{Physical Review B}, vol. 52, no. 6 (1995): 4477. DOI: 10.1103/PhysRevB.52.4477.

\bibitem{Pambianchi:96}
M. S. Pambianchi, L. Chen, and S. M. Anlage, ``Complex conductivity of proximity-superconducting Nb/Cu bilayers,'' \textit{Physical Review B}, vol. 54, no. 5 (1996): 3508. DOI: 10.1103/PhysRevB.54.3508.

\bibitem{Wangr:00}
R. F. Wang, S. P. Zhao, G. H. Chen, and Q. S. Yang, ``Magnetic penetration depth in Nb/Al and Nb/Cu superconducting proximity bilayers,'' \textit{Physical Review B}, vol. 62, no. 17 (2000): 11793. DOI: 10.1103/PhysRevB.54.3508.


\end{thebibliography}
\end{document}